\documentclass[aps,pre,twocolumn,groupedaddress,showpacs,floatfix]{revtex4}
\usepackage{psfrag}
\usepackage{graphicx}
\usepackage{pst-all}
\usepackage{color}
\usepackage{amsmath}
\usepackage{graphicx,psfrag}
\usepackage{epsfig}
\usepackage{graphics}
\newcommand{\comment}[1]{}

\def\reacts#1{\mathop{\longrightarrow}\limits^{\raisebox{0.4ex}{$\scriptscriptstyle {#1}$}}}

\begin{document}
\title{Nonequilibrium critical behavior of a species coexistence model}
\author{Heiko Reinhardt$^1$, Frank B\"ohm$^1$, Barbara Drossel$^1$, and Haye Hinrichsen$^2$}
\affiliation{$^1$ Institut f\"ur Festk\"orperphysik,  TU Darmstadt,
Hochschulstra\ss e 6, 64289 Darmstadt, Germany }
\affiliation{$^2$ Fakult\"at f\"ur Physik und Astronomie, 
         Universit\"at W\"urzburg, Am Hubland, 97074 W\"urzburg, Germany}
\date{\today}
\begin{abstract}
A biologically motivated model for spatio-temporal coexistence of two competing species is studied by mean-field theory and numerical simulations. In $d \geq 2$ dimensions the phase diagram displays an extended region where both species coexist, bounded by two second-order phase transition lines belonging to the directed percolation universality class. The two transition lines meet in a multicritical point, where a non-trivial critical behavior is observed.
\end{abstract}
\maketitle
\parskip 2mm

\section{Introduction}
\label{sec:intro}

Nonequilibrium models with phase transitions into absorbing states
arise in studies of biological populations, chemical reactions,
spreading of diseases, and many
other~\cite{MarroDickman,Kinzel85,Hinrichsen00,OdorReview,Luebeck04}.
An absorbing phase is a subspace of states that can be reached but not
be left by the dynamics. In biologically motivated models absorbing
phases occur whenever a system reaches a configuration or a dynamical
state from where it cannot escape. In many cases an absorbing state is
characterized by a complete loss of activity.  It is believed that
phase transitions into an absorbing state generically belong to the
directed percolation (DP) universality
class~\cite{Janssen81,Grassberger82,jan05}. DP is extremely robust and
includes even models with several species of
particles~\cite{ZGB86,GLB89,JFD90} and infinitely many absorbing
states~\cite{Jensen93,MGDL96}.  Exceptions from DP are usually
observed in models with long-range interactions, quenched randomness,
and non-conventional symmetries such as a $Z_2$-symmetry or
conservation of particle number.

In models with several species of particles and several absorbing
states, there exist in addition to transitions where one species goes
extinct, multicritical points where several reduced control parameters vanish
simultaneously. Such multicritical points are characterized by a
hierarchy of order parameter exponents with only one of them being the
DP exponent, and by a crossover exponent $\phi$~\cite{tau98,gol99}.
Again, the multicritical behavior is different in the presence of
symmetries~\cite{bas96}. 

Most DP models have in common that active sites may become inactive at a certain rate, i.e.
particles can disappear spontaneously. But even models with a different
mechanism for particle removal may belong to the DP class. For example,
in branching and annihilating random walks particles can only annihilate
pairwise, leading to an algebraic decay in the subcritical phase. 
Until recently, it was believed that such models
do not show a DP-like phase transition in dimensions larger than two. 
However, as has been shown by Canet \textit{et al}~\cite{can04}, 
in higher dimensions such transitions occur also, but
only beyond a minimum annihilation rate, so that freshly produced
offspring can annihilate with its parent with a sufficiently large
probability before diffusing away.

In this article, we consider a two-species model, which is inspired by
biology. A special feature of this model is that one of the two
species (denoted as $S$) cannot vanish spontaneously, instead it can
only be destroyed by the other one (called $A$).  In one dimension it
turns out that the two species cannot coexist and the phase transition
is discontinuous for the first species. In higher dimensions the two
species can coexist in a certain region of the phase diagram, giving
rise to three different phases.  We find numerically that all phase
boundaries where one species becomes extinct, are in the DP class,
while the multicritical point where the two critical lines meet, shows
a more complex scaling behavior. Near the multicritical point, the
critical region becomes very small, and therefore the DP behavior can
hardly be seen in numerical simulations.  We apply numerical studies
as well as mean-field calculations and scaling arguments. In addition
to the stationary state, we also investigate the dynamical behavior
starting from a random initial state. We find that two different
scaling forms are needed to describe different dynamical regimes.

The outline of this paper is as follows: In the next section, we
motivate and define our model. In Section~\ref{sec:mf} we formulate a
mean field theory and discuss the corresponding phase diagram.
Numerical simulations in one and two spatial dimensions are presented
in Section~\ref{sec:numerics}. In Section~\ref{sec:scaling} we suggest
phenomenological scaling theory for the dynamical and the stationary
regimes. Finally, we summarize our findings in
Section~\ref{sec:conclusions}.

\section{Definition and motivation of the model}
\label{sec:model}

The model is defined on a $d$-dimensional hypercubic lattice with $L^d$ sites. 
Each lattice site is occupied either by species $A$, or by species $S$, 
or it is empty. The system evolves  random-sequentially 
according to the following dynamic rules: 

\begin{itemize}
\item Any site that has a nearest neighbor occupied by species $A$
  becomes itself occupied by species $A$ with rate $\lambda_A$,
  irrespective of its previous state.
\item An empty site that has a nearest neighbor occupied by species
  $S$ becomes itself occupied by species $S$ with rate $\lambda_S$.
\item A site occupied by species $A$ turns into an empty site with 
rate $\mu$. 
\end{itemize}

\noindent
This means that the model follows the reaction-diffusion scheme
\begin{eqnarray}
&&A+\emptyset \reacts{\lambda_A} 2A,  \qquad A \reacts{\mu}  \emptyset,\\[2mm]
&&A+S\reacts{\lambda_A}  2A, \qquad S+\emptyset \reacts{\lambda_S}  2S.\nonumber
\end{eqnarray}
The model can be considered as a patch occupancy model for the
coexistence of two consumer species feeding on the same resources.
The two species cannot coexist in the same patch, and therefore each
patch is occupied at most by one of the two species~\cite{spatialecology}. 
Species $A$ has the higher fitness and therefore displaces species $S$ when it enters a
patch occupied by $S$. However, species $A$ has a certain risk of
dying out in a patch, for instance because it cannot respond well to
adverse circumstances, such as illness or extreme weather conditions.
$A$ and $S$ may stand for an asexual and a sexual species, for
instance oribatid mites in the soil~\cite{pal92}, or ostracods in
ponds lakes~\cite{scho98}, where asexual and sexual species are known
to have coexisted for a long time. Asexual species have a higher
reproduction rate, but can accumulate deleterious mutations~\cite{mul64}, 
which reduce their fitness below that of the sexual
variant.

In our model the dynamics of species $A$ is not affected by the presence of $S$
since empty sites and sites occupied by $S$ can both be conquered by
$A$. The dynamics of species $A$ is therefore a version of DP, and the
phase transition of species $A$ to extinction is in the DP
universality class. More specifically, the dynamic rules of the $A$'s
are equivalent to those of a so-called contact process~\cite{Harris74}, which
is a well-studied lattice model of DP with random-sequential updates.
Precise estimates of the percolation threshold 
$\lambda_A^{(c)}/\mu$~\cite{JensenDickman93,Dickman99,LuebeckWillmann05} and the  
critical exponents~\cite{Luebeck04} can be found 
in the literature and are summarized in Table~\ref{CPData}.

\begin{table}
\begin{tabular}{|c|c|c|c|c|}
\hline \,\,$d$\,\, 
                & $\lambda_A^{(c)}/\mu $ &  $\beta$	     & $\nu_\perp$    & $\nu_\parallel$ \\  \hline
\hline $1$      & $1.648924(11)$         &  $0.276486(8)$ & $1.096854(4)$  & $1.733847(6)$ \\ 
\hline $2$      & $0.41219(1)$           &  $0.583(3)$    & $0.733(8)$     & $1.295(6)$ \\ 
\hline $3$      & $0.219477(2)$          &  $0.813(9)$    & $0.584(5)$     & $1.110(1)$ \\ 
\hline $4$      & $0.14938(2)$           &  $1$           & $1/2$          & $1$ \\ 
\hline 
\end{tabular}
\caption{Critical percolation thresholds $\lambda_A^{(c)}/\mu$ of the contact process
of $A$-particles and the corresponding critical exponents of DP in $d$+1 dimensions.
\label{CPData}}
\end{table}

In absence of species $A$, species $S$ spreads and eventually occupies
the entire system. In presence of species~$A$ patches of species $S$
can only be destroyed if they are invaded by species $A$. Since the
distribution of $A$ is not homogeneous but highly correlated, 
patches occupied by $S$ are not destroyed randomly, and it is not 
obvious that the phase transition to extinction of $S$ has to be 
in the universality class of directed percolation. 
We will see below that this is nevertheless the case 
in dimensions larger than one, although the DP behavior is
difficult to see when the density of $A$ is low.

\section{Mean field theory}
\label{sec:mf}

Let us first discuss the mean-field theory of the model. 
The mean-field equations for the densities of species $A$ and $S$ are
\begin{eqnarray}
\label{MF-Gl1}  
\dot{\rho}_A(t) &=& \lambda_A\rho_A(t)\bigl[1-\rho_A(t)\bigr] - {\mu}\rho_A(t)\,,\\
\label{MF-Gl2}  
\dot{\rho}_S(t) &=& \lambda_S\rho_S(t)\bigl[1-\rho_A(t)-\rho_S(t)\bigr]-
                      \lambda_A\rho_S(t)\rho_A(t).\nonumber\\
\end{eqnarray}
The mean-field approximation can be viewed as the case where the $A$
or $S$ can move to any patch of the lattice with a given rate
$\lambda_A/L^d$ or $\lambda_S/L^d$, and not only to nearest neighbors,
in the limit of infinite system size $L\to \infty$.  

\subsection{Stationary case}

In order to determine the mean-field phase diagram we computed
the stationary solutions of Eqs.~(\ref{MF-Gl1})-(\ref{MF-Gl2}).
Table~\ref{FP-MF} lists the fixed points and the corresponding
stationary densities $\rho_A$ and $\rho_S$ 
together with the conditions for their stability. The fixed
point $F_1$ is unstable since $\lambda_S$ and $\lambda_A$ cannot be
negative. The other three fixed points are stable in different parts of
parameter space. 
\begin{table}
\begin{tabular}{|c|c|c|c|c|c|}\hline\
Fixed point & $\rho_A$ & $\rho_S$ &\multicolumn{2}{|c|}{conditions for stability}\\\hline\hline
$F_1$ &0&0& $\lambda_S<0$ &$\lambda_A < \mu$ \\\hline
$F_2$ &0 & 1& $\lambda_S>0$ &$\lambda_A<\mu$ \\\hline
$F_3$ & $1-\frac{\mu}{\lambda_A} $ & 0& $\lambda_S < \frac
 {\lambda_A^2}{\mu}-\lambda_A $& $\lambda_A>\mu$ \\ \hline
$F_4$ & $1-\frac{\mu}{\lambda_A} $ &  $\frac{\mu}{\lambda_A}
 +\frac{\mu-\lambda_A}{\lambda_S} $& $\lambda_S > \frac{\lambda_A^2}{\mu} - \lambda_A$ & $\lambda_A>\mu$ \\\hline
\end{tabular}
\caption{\label{FP-MF}Fixed points of the mean-field equations (\ref{MF-Gl1})-(\ref{MF-Gl2}) and the corresponding conditions for stability.}
\end{table}

\begin{figure}
\begin{center}
\includegraphics[scale=0.33, angle=270] {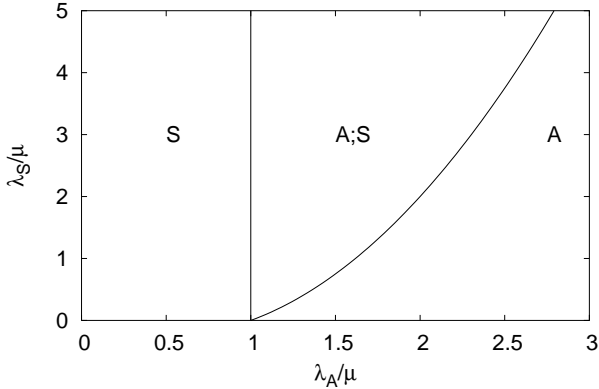}
\end{center}
\caption{Mean-field phase diagram. In phase \textbf{S} species $A$ dies out so that species $S$ conquers the whole system, while in phase \textbf{A} species $S$ dies out. In phase \textbf{A;S} both species coexist, leading to a non-trivial fluctuating stationary state.}
\label{MF-plot}
\end{figure}

In the phase diagram the three fixed points correspond to three 
different phases (see Fig.~\ref{MF-plot}). In the subcritical phase
(marked by \textbf{S}) species $A$ dies out so that species $S$ eventually
conquers the whole system, approaching the stationary density $\rho_S=1$.
In the coexistence phase (marked by \textbf{A;S}) both species can coexist,
leading to a non-trivial fluctuating stationary state with the densities
\begin{eqnarray}
\label{StationaryA}
\rho_A &=& 1-\frac{\mu}{\lambda_A}\,, \\
\label{StationaryS}
\rho_S &=& \frac{\mu}{\lambda_A}+\frac{\mu-\lambda_A}{\lambda_S}\, .
\end{eqnarray}
Finally, in the phase denoted by \textbf{A} in Fig.~\ref{MF-plot}, the density
of species $A$ is so high that species $S$ becomes extinct. Here Eq.~(\ref{StationaryA})
is still valid while $\rho_S=0$. 

The three phases are separated by two lines of continuous phase transitions,
a vertical one at $\lambda_A/\mu=1$ and a curved one in form of a segment of a parabola
\begin{equation}
\label{MFCriticalLine}
\lambda_S^*=\lambda_A^*(\frac{\lambda_A^*}{\mu}-1) \,.
\end{equation}
Approaching the vertical line from right to left the stationary density $\rho_A$
vanishes linearly. The critical exponent $\beta$ is therefore 1, 
just as in the standard mean-field theory of directed percolation. The same applies
to the density $\rho_S$ approaching the curved phase transition line from the left.
This supports our numerical findings (see below) that both lines represent DP transitions.
 
The two phase transition lines  meet at the multicritical point
$\lambda_A/\mu=1$; $\lambda_S=0$. In order to discuss the behavior
in the neighborhood of the multicritical point in more detail, 
it is convenient to
introduce the parameters 
\begin{equation}
\label{epsilonmf}
\epsilon_A = \lambda_A/\mu-1\,, \qquad \epsilon_S = \lambda_S/\mu\,,
\end{equation}
which measure the reduced distances from the multicritical point in
horizontal and vertical direction, respectively. In terms of these
parameters the curved transition line~(\ref{MFCriticalLine}) is given by
\begin{equation}
\label{CurvedLineEpsilon}
\epsilon_S^* = \epsilon_A^* ( 1+\epsilon_A^*)\,.
\end{equation}
Close to the multicritical point, where both parameters are small, the densities
scale as
\begin{eqnarray} 
\label{MFA}
\rho_A &=& 1-\frac 1 {1+\epsilon_A} \simeq \epsilon_A\,, \\
\label{MFS}
\rho_S &=& \frac 1 {1+\epsilon_A} - \frac{\epsilon_A}{\epsilon_S}
\simeq 1-\frac{\epsilon_A}{\epsilon_S} \, .\label{rhos}
\end{eqnarray}
As we decrease  $\epsilon_A$ from $\epsilon_S$ to 0, we move
horizontally from the curved to the vertical transition line, and
$\rho_S$ increases linearly from 0 to 1. 
Similarly, as we increase $\epsilon_S$ from $\epsilon_A$ to a value 
much larger than $\epsilon_A$, we move from the curved transition line
vertically upwards, and $\rho_S$ increases from 0 to a value close to 1. 
Denoting by 
\begin{equation}
\Delta_A=\epsilon_A^*-\epsilon_A\,\hbox{ and }\, \Delta_S=\epsilon_S-\epsilon_S^*
\end{equation}
the horizontal and vertical distances between a given point $(\epsilon_A,\epsilon_S)$ and  the curved transition line, 
we have for $\Delta_A\ll\epsilon_A, \Delta_S \ll\epsilon_S$
\begin{equation}
\label{MFCrossover}
\rho_S = \frac {\Delta_A}{\epsilon_S} = \frac{\Delta_S}{\epsilon_A}.
\end{equation}
Therefore, the linear decrease of $\rho_S$ does not depend on the direction
in which the curved phase transition line is approached. However,
the \textit{slope} increases and finally diverges as we approach the
multicritical point. This indicates that the transition with respect to $S$
\textit{at} the multicritical point may no longer belong to the DP class.

\subsection{Time-dependent mean field solution}
\label{mfdyn}
%
In order to understand the dynamical properties, let us now 
solve the time-dependent mean field equations~(\ref{MF-Gl1})-(\ref{MF-Gl2}) explicitely. 
Since species $A$ evolves independently according to the rules of DP 
Eq.~(\ref{MF-Gl1}) is autonomous. With $\epsilon_A = \lambda_A/\mu-1$ and  
$\epsilon_S = \lambda_S/\mu$ the solution (up to a shift in time) reads
\begin{equation}
\label{Asolution}
\rho_A(t) = \frac{\epsilon_A}{(1+\epsilon_A)\,\bigl[1-\exp(-\epsilon_A \mu t)\bigr]}\,. 
\end{equation}
Close to criticality in the active phase $0<\epsilon_A \ll 1$ this density first decays as a power law 
\begin{equation}
\label{AlgebraicDecay}
\rho_A(t) \simeq \frac{1}{(1+\epsilon_A)\,\mu t}
\end{equation}
until it crosses over to the stationary value given in Eq.~(\ref{MFA}) 
at some typical crossover time $t_c \sim \frac{1}{\mu \epsilon_A}$.

Inserting the solution~(\ref{Asolution}) into the second mean field equation (\ref{MF-Gl2}), it is in principle possible to compute the time-dependent density $\rho_S(t)$. 

\begin{figure}
\begin{center}
\includegraphics[width=70mm] {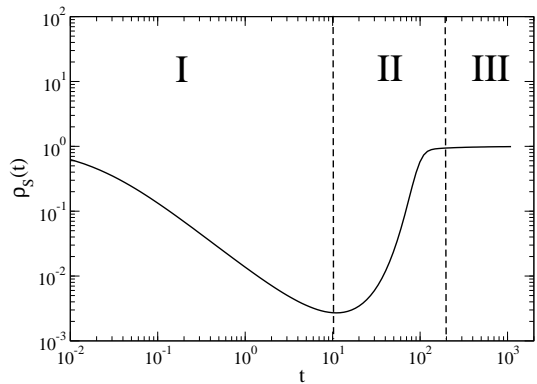}
\end{center}
\vglue -6mm
\caption{Temporal evolution of the density $\rho_S(t)$ according to the mean-field solution~(\ref{ExactMFSolution}) for $\epsilon_A=0$, $\epsilon_S=0.1$, $\mu=1$ and $c=10^3$. The three regimes are explained in the text.}
\label{fig:mfdyn}
\end{figure}

Let us first consider the case $\epsilon_A=0$ (the vertical line in Fig.~\ref{MF-plot}), 
where the DP process of species $A$ is critical. Inserting Eq.~(\ref{AlgebraicDecay}) 
into Eq.~(\ref{MF-Gl2}), the second mean field equation reduces to
\begin{equation}
\label{FormalDifferentialEquation}
\dot{\rho}_S(t) \;=\; -\frac{\rho_S(t)}{t}\,\Bigl[ 1 + \epsilon_S + \epsilon_S \mu t\, \bigl(\rho_S(t)-1\bigr) \Bigr]\,.
\end{equation}
At the multicritical point $\epsilon_S=\epsilon_A=0$  this differential equation further reduces to $\dot{\rho}_S(t) =-t^{-1}\rho_S(t)$ so that the density $\rho_S(t) \sim 1/t$ decays in the same way as $\rho_A(t)$. For $\epsilon_S>0$, however, we find the formal solution
\begin{equation}
\label{ExactMFSolution}
\rho_S(t) =
\frac{\exp(\epsilon_S\mu t)}
{(\epsilon_s \mu t)^{1+\epsilon_S} ´
\left[ C + \int_1^{\epsilon_S \mu t} e^z \, z^{-1-\epsilon_s} \, {\rm d} z \right]},
\end{equation}
where $C$ is an integration constant. To get an impression about the behavior for $\epsilon_S>0$, we plotted this function in Fig.~\ref{fig:mfdyn}. As can be seen, there are three different dynamical regimes enumerated by I, II, and III. In the first regime the density $\rho_S(t)$ decreases algebraically as $1/t^{1+\epsilon_S}$, followed by a quick increase in the second regime until the density saturates at 1 in regime III.

The observed behavior can be interpreted as follows. Starting with a homogeneously distributed mixture of species $A$ and $S$, the displacement of $S$ by $A$ initially dominates the dynamics. In regime II the spreading of $S$ to empty patches, as expressed by the linear term $\lambda_S\rho_S(t)$ on the r.h.s. of Eq.~(\ref{MF-Gl2}), becomes relevant so that the $S$-population begins to grow exponentially in the voids of the critical $A$-process. The growth of the $S$-population continues until all accessible voids are filled and the density $\rho_S(t)$ reaches its stationary maximum in regime III.

Next, let us consider the dynamics on the curved phase transition line close to the multicritical point. Using Eqs.~(\ref{CurvedLineEpsilon}) and (\ref{Asolution}), expanding $\epsilon_S$ and $\rho_A$ in powers of $\epsilon_A$ and keeping only terms up to the third order in the densities and/or the control parameter, Eq.~(\ref{MF-Gl2}) turns into
\begin{equation}
\frac{1}{\mu}\,\dot \rho_S(t) = -\frac{\epsilon_A(1+\epsilon_A)}{\exp(\epsilon_A\mu t)-1}\rho_S(t)
-\epsilon_A\rho_S^2(t)\, .
\end{equation}
The solution is
\begin{equation}
\rho_S(t) = 
\frac{[1-\exp(-\epsilon_A \mu t)]^{-1-\epsilon_A}}
{\tilde{C}+\int_1^{\epsilon_A \mu t} (1-e^{-z})^{-1-\epsilon_A}\,{\rm d}z}\, 
\end{equation}
with an integration constant $\tilde{C}$.  This is a monotonous decay of the
density $\rho_S$. For short times, it is approximated by $\rho_S \sim
1/t$, as we have obtained before. For large times, the density
$\rho_S$ approaches the behavior $\rho_S \simeq 1/\epsilon_A \mu t$, i.e.,
the decay is delayed. This large-time behavior is
independent of the initial value of $\rho_A$, and it is therefore also
obtained when would have started with the stationary solution for $\rho_A$.

\section{Numerical simulations}
\label{sec:numerics}

\subsection{Simulations in 1+1 dimensions}
\label{oned}

\begin{figure}
\begin{center}
\psfrag{t}{\large t}
\psfrag{x}{\large x}
\includegraphics[scale=0.2, angle=0] {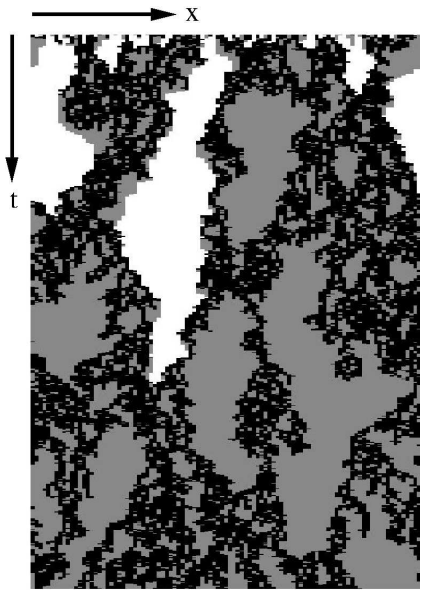}
\end{center}
\caption{Space-time plot of the 1+1-dimensional
  model. White and black sites correspond to species $S$ and $A$
  respectively, and  grey sites correspond to empty patches.}
\label{fig1}
\end{figure}

The 1+1-dimensional case is special in so far as
species $A$ and species $S$ cannot coexist since
the world lines of $A$ and $S$ cannot cross each other. Hence a species 
that persists forever leaves no room for a world line of
the other species. Fig.~\ref{fig1} shows a space-time plot of $A$,
$S$ and empty sites in the parameter range where $A$
persists. As can be seen, starting from a random initial state, 
species $S$ eventually dies out. 

\begin{figure}
\begin{center}
\includegraphics[scale=0.34, angle=270] {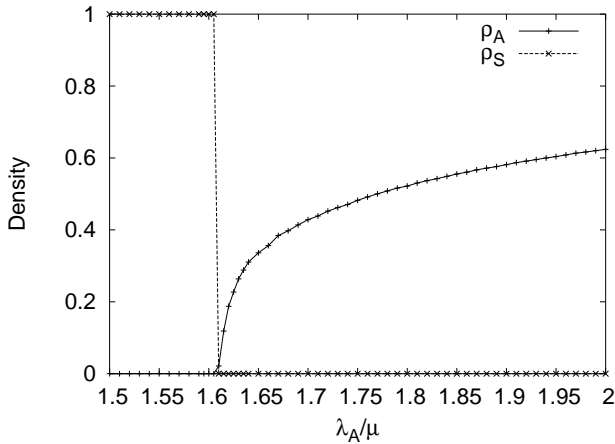}
\end{center}
\caption{Stationary density as a function of the control parameter $\lambda/\mu$ in one dimension (here $\lambda=\lambda_A=\lambda_S$).}
\label{fig2}
\end{figure}

The stationary densities for $A$ and $S$ as functions
of the control parameter $\lambda_A/\mu$ were measured in
numerical simluations using a system size of $L=50\,000$ sites.  
As shown in Fig.~\ref{fig2}, species $S$ undergoes a first-order 
phase transition from unit density to zero density at the critical 
point $\lambda_A^{(c)}$ of species $A$. 
 
\subsection{Stationary properties in 2+1 dimensions}
\label{twod}

\begin{figure}
\begin{center}
\includegraphics[scale=0.34, angle=270] {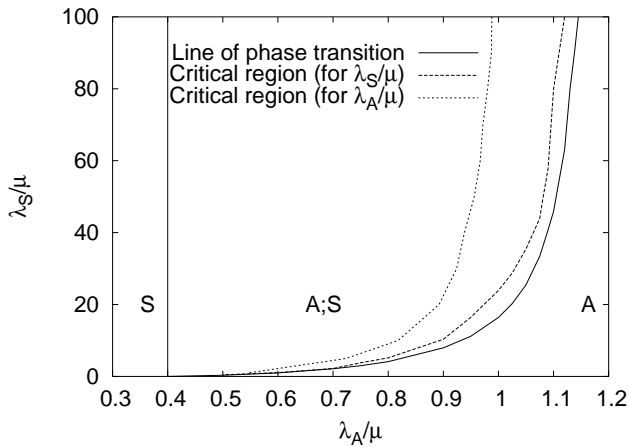}
\end{center}
\caption{Numerically determined phase diagram in 2+1 dimensions.
The dotted lines denote the size of the critical regions for
variation of $\lambda_A/\mu$(dotted line) and $\lambda_S/\mu$ (dashed line).}\label{krit}
\end{figure}

In higher spatial dimensions both species can coexist and one obtains a phase diagram
with two continuous transition lines, which is qualitatively similar to that of the mean-field approximation.
The numerically determined phase diagram in 2+1 dimensions is shown in Fig.~\ref{krit}. 
The critical thresholds obtained by varying $\lambda_A/\mu$ agrees with those obtained 
by varying $\lambda_S/\mu$, which confirms the correctness of the results.  
The multicritical point is located at
\begin{equation}
\label{multicritical}
\lambda_A^{(c)}/\mu=0.41219(1)\,,\quad
\lambda_S^{(c)}=0\,,
\end{equation}
where the evolution of species $A$ is critical and branching of $S$ 
is not allowed. In fact, even for very small $\lambda_S>0$
species $S$ should be able to survive as species $A$ becomes extinct, 
eventually approaching $\rho_S=1$. Therefore, as in mean-field theory, 
the multicritical point is located exactly at $\lambda_S^{(c)}=0$. 

In order to determine the critical exponents in the stationary state, 
we performed simulations at several points in the phase diagram
close to the phase boundaries. The lattice size was $L^2=300\times300$ sites. 
After $200\,000$ time steps the densities of $A$ and $S$ reached a 
stationary value. The fluctuations were within 5\% of
this value. To determine the mean densities, the density was averaged
over an additional $100\,000$ time steps.  

\begin{figure}
\begin{center}
\includegraphics[scale=0.3, angle=270] {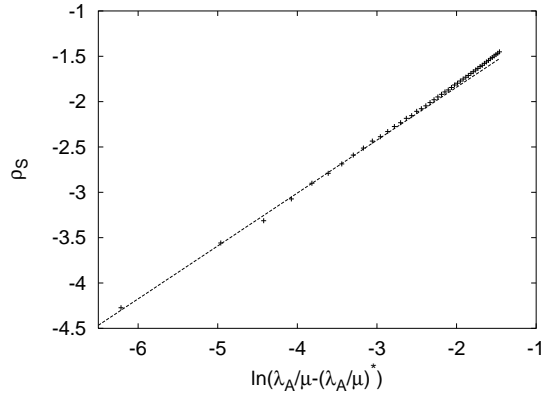}
\end{center}
\caption{Logarithmic plot of the stationary density $\rho_S$ for $\lambda_S/\mu=60$ 
    as $\lambda_A/\mu$ is varied. A power-law fit in the
    interval $|\lambda_A/\mu-(\lambda_A/\mu)|^*<=0.05$ yields the estimate
    $\beta_S=0.583(3)$.}
\label{log-a}
\end{figure}

For the extinction of species $A$ at the vertical transition line, the 
expected directed percolation behavior $\rho_A \sim (\lambda_A-\lambda_A^{(c)})^{\beta_A}$ 
with $\beta_A = \beta_{DP} \simeq 0.583$ is clearly seen in the simulations.  
However, the evaluation of the critical exponent $\beta_S$ in the vicinity of the curved
phase transition line turned out to be more difficult. For example, 
close to the multicritical point the estimates 
were found to depend on the direction in which the line is approached, 
which is known to be impossible in standard critical phenomena. In fact,
reliable estimates could only be obtained far away from the multicritical point. 
Choosing $\lambda_S/\mu = 60$, where the density of species $A$ is high, and estimating
the exponent only in a limited range of the data close to criticality (see Fig.~\ref{log-a}),
we find numerical evidence that the transition along the curved line
belongs to the DP universality class.

\begin{figure}
\begin{center}
\includegraphics[scale=0.3, angle=270] {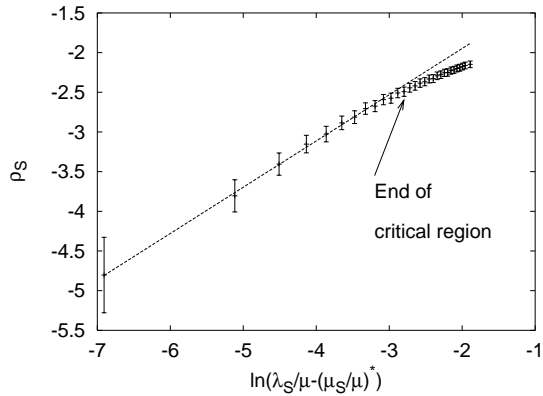}
\end{center}
\caption{Logarithmic plot of $\rho_S$, including 
    its fluctuations for $\lambda_A/\mu=1$. The power-law fit gives
    the exponent $\beta=0.584(9)$.}
\label{log-plot}
\end{figure}

Approaching the multicritical point deviations from the expected power law
behavior become more pronounced. This is demonstrated in Fig.~\ref{log-plot}, 
where we measured $\rho_S$ for $\lambda_A/\mu=1$, approaching the
curved transition line vertically from above. To estimate 
the size of the critical region, the data points for $\rho_S$
were plotted including error bars. Then, a power-law was fitted to the
first $k$ data points, with $k$ being small enough to be close to
the scaling regime. The measured critical exponent was compatible with 
$\beta_S = 0.585 \pm 0.05$ everywhere. This confirms that the entire  
phase transition line (the solid curved line in Fig.~\ref{krit}) except
for the multicritical point belongs to the DP universality class.

In order to quantify how the critical behavior is approached we
defined critical regions whose size is determined by the requirement
that the error bar of the 
last data point in Fig.~\ref{log-plot} still touches the power-law fit. 
The size of these regions with respect to variations of $\lambda_A$
and $\lambda_S$ are indicated in Fig.~\ref{krit} by dashed lines.
Although their size is arbitrary in the sense that
it depends on the chosen accuracy of the computer simulations, 
it illustrates why estimates far away from the multicritical point 
are more reliable. Moreoever, it is interesting to note that the critical region for 
variation of $\lambda_A/\mu$ is much larger than the one for $\lambda_S/\mu$.

As we approach the multicritical point the critical regions become
narrower for both variations. In the immediate vicinity of the
multicritical point ($\lambda_A/\mu<0.5$ and $\lambda_S/\mu<1$), a
reliable estimation of the critical exponent $\beta_S$ becomes
extremely difficult due to huge fluctuations of $\rho_S$.  Here we
have an almost instant increase in the density of $S$ as we vary the
control parameters. Below, we will determine the shape of the curved
phase transition line close to the multicritical point with a better
precision by using dynamical simulations.

\subsection{Dynamical properties in 2+1 dimensions}
\label{twoddyn}

In order to compare the time dependence of $\rho_S(t)$ in 2+1
dimensions with the mean field prediction shown in
Fig.~\ref{fig:mfdyn}, we first performed numerical simulations along
the vertical line $\lambda_A=\lambda_A^{(c)}$, where the DP process of
species $A$ is critical. Starting with a homogeneous random mixture of
both species, we monitor the evolution of the densities $\rho_A(t)$
and $\rho_S(t)$ for various values of $\lambda_S/\mu$.

\begin{figure}
\begin{center}
\includegraphics[width=72mm] {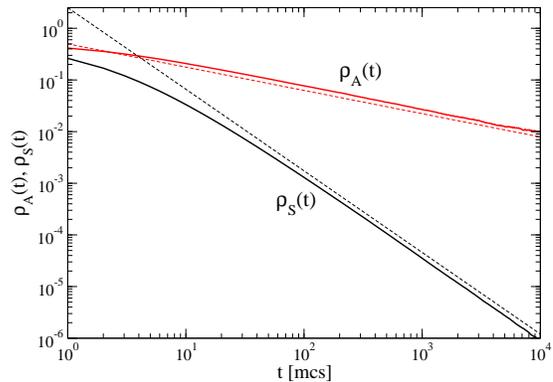}
\end{center}
\vglue -6mm
\caption{Decay of the densities $\rho_A(t)$ and $\rho_S(t)$ at the
multicritical point in 2+1 dimensions. The dashed line indicate the
slopes $-\delta$ and $-\theta$ (see text).}
\label{fig:persist}
\end{figure}  

The results of simulations {\it at} the multicritical point $\lambda_A=\lambda_A^{(c)}, \lambda_S=0$
are shown in Fig.~\ref{fig:persist}. As expected,  the density of species $A$ decays as
\begin{equation}
\rho_A(t) \sim t^{-\delta}\,,
\end{equation}
where $\delta=\beta/\nu_\parallel \approx 0.451$ is the
usual decay exponent of DP in two dimensions. 
Similarly, the density of species~$S$ is found to decay algebraically
as $\rho_S(t)\sim t^{-\theta}$ with an exponent $\theta \approx 1.6$. 

The value of the exponent $\theta$ can be explained
as follows: Initially each site is randomly occupied by $A$ or $S$ with equal probability. 
As time evolves the critical DP-process of species $A$ gradually removes species $S$, 
leading to a monotonous decrease of $\rho_S(t)$. As species $S$ does not create offspring
at the multicritical point, 
all sites occupied by $S$ at time $t$ are precisely those sites that have never been 
occupied before by species $A$. Hence $\rho_S(t)$ is the probability that a given 
site has never been visited by the DP process of species $A$. In the literature 
this probability is known as the \textit{local persistence probability}~\cite{HinrichsenKoduvely98} 
$P_\ell(t)$ of directed percolation. This quantity was shown to decay algebraically as 
$P_\ell(t)\sim t^{-\theta}$, where $\theta$ is the so-called local persistence exponent 
which seems to be independent of the other DP exponents. In one spatial dimension its 
value was estimated by $\theta=1.50(1)$, while in two dimensions we find a slightly 
higher value $\theta=1.58(3)$. This observation allows us to conclude that $\rho_S(t)$ 
decays at the multicritical point as
\begin{equation}
\label{Persistence}
\rho_S(t) \sim t^{-\theta}\,. \qquad (\epsilon_A=\epsilon_S=0)
\end{equation}
Increasing $\lambda_S$ (while keeping the $A$-process critical) we allow species $S$
to expand in the voids of the $A$-process. 
In this case the algebraic decay (\ref{Persistence}) is expected to 
cross over to a growth of surviving $S$-domains.
This density continues to increase until the entire 
empty space is conquered so that
$\rho_S$ approaches the value $1$. Our simulation results for this process are shown in the top left panel of Figure \ref{fig:scaling}.
Obviously, the temporal behavior of $\rho_S(t)$ resembles the mean field solution in Fig.~\ref{fig:mfdyn}, 
where three different dynamical regimes were identified. However, as an important difference we note
that in regime II the density $\rho_S$ increases \textit{algebraically} as $t^d$ in finite-dimensional systems
whereas in the mean-field limit the increase is exponential. Consequently, the two crossover time scales
between the regimes are expected to scale differently.

Next, we performed numerical simulations on the curved phase
transition line. The results are shown in the top right panel of
Figure \ref{fig:scaling}. For short times, $\rho_S$ decays again as
$t^{-\theta}$, i.e., the difference between the two critical lines is
not yet felt. For large times, when $\rho_A$ has become stationary,
$\rho_S$ decays as $t^{-\delta}$, i.e., it behaves as a directed
percolation process at its critical point. Between these two regimes,
there is an increase in the density $\rho_S$, which is not seen in the
mean-field theory. As we will discuss further below, the reason for
this is that the characteristic time scales for the cutoff of the
initial decay and the onset of the DP decay scale differently with $\epsilon_S$, 
while there is no distinction between the two scales in mean-field theory. 
Therefore, in two- or three-dimensional systems the density $\rho_S$ increases in the
intermediate regime, where the spreading to empty patches has become
relevant, while the critical DP behavior, which requires stationary of
$\rho_A$, is not yet established. In analogy to the behavior on the
vertical transition line, we denote the three regimes of initial
decay, intermediate increase and ultimate decay with regime I, II, and
III.

\subsection{Functional form of the curved transition line}
\label{PhaseTransitionLine}
%
%
In the mean-field case the curved transition line was shown to be a segment of a parabola which terminates at the multicritical point with slope $1$ (see Eq.~(\ref{MFCriticalLine})). In low dimensions it is therefore natural to assume a power-law of the form
\begin{equation}
\label{TransitionLine}
\epsilon_S^* \propto (\epsilon_A^*)^y\,,
\end{equation}
where again
\begin{equation}
\epsilon_A = (\lambda_A-\lambda_A^{(c)})/\mu\,, \qquad \epsilon_S = \lambda_S/\mu\,
\end{equation}
are the reduced parameters and $y$ is an unknown exponent with the mean-field value $y^{\begin{tiny}\rm MF\end{tiny}}=1$. As in the mean-field case, this power law may be superposed by next-leading algebraic corrections.

\begin{figure}
\begin{center}
\includegraphics*[width=85mm] {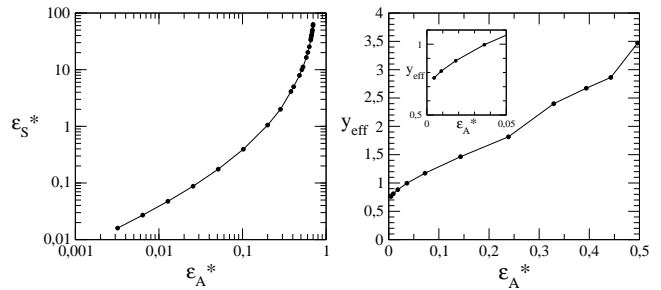}
\end{center}
\vglue -5mm
\caption{Left panel: Critical line for the extinction of species $S$ 
for small $\epsilon_A^*$ and $\epsilon_S^*$. Right panel: 
Corresponding local slopes for a visual extrapolation. The inset
suggests an asymptotic value $y \approx 0.7$.
\label{kritlin}}
\end{figure}

In order to determine the value of the unknown exponent $y$ in 2+1 dimensions 
numerically, we adjusted the parameters 
$\epsilon_A$ and $\epsilon_S$ in such a way that $\rho_S(t)$
decays eventually in the same way as in a DP process. However, the determination of $y$ turns
out to be extremely difficult. In particular, the estimates seem to drift to 
smaller values as we approach the multicritical point. 
Extrapolating the local slopes (see Fig.~\ref{kritlin}) we arrive at the conclusion that $y \leq 1$,
probably $y \approx 0.7(2)$. 

%

\section{Scaling properties}
\label{sec:scaling}

In this section, we interpret the observed results in terms of phenomenological scaling arguments. It turns out that the presence of different scales makes it impossible to formulate a unified scaling theory for the whole dynamical range, instead we find two different scaling forms for regimes I,II and II,III, respectively with a common overlap in region II. 

Starting point is the DP-process of species $A$ which decouples from the dynamics of species $S$. Close to criticality this DP process is known to be invariant under scale transformations 
\begin{equation}
\epsilon_A \to \Lambda \epsilon_A \,, \quad
\rho_A \to \Lambda^{\beta}\rho_A \,,\quad
t \to \Lambda^{-\nu_\parallel} t \,, 
\end{equation}
where $\Lambda>0$ is a scaling factor while $\beta$ and $\nu_\parallel$ are DP exponents listed in Table~\ref{CPData}. As usual scale invariance implies that the density is a homogeneous function, i.e.,
\begin{equation}
\rho_A(\epsilon_A,t)=
\Lambda^{-\beta}\rho_A(\Lambda \epsilon_A, \Lambda^{-\nu_\parallel} t)\,.
\end{equation}
Setting $\Lambda=t^{1/\nu_\parallel}$ and suppressing non-universal metric factors one obtains the usual scaling form
\begin{equation}
\rho_A(t)=t^{-\delta} \, F_A(\epsilon_A t^{1/\nu_\parallel})\,,
\end{equation}
where $F_A$ is a universal scaling function. In order to derive a similar scaling form for the dynamics of species~$S$ in the vicinity of the multicritical point, it is natural to postulate analogous scaling properties for $\epsilon_S$ and $\rho_S$, i.e.
\begin{equation}
\label{PostulatedScaling}
\rho_S\to \Lambda^{x_1}\rho_S\,, \qquad
\epsilon_S \to \Lambda^{x_2}\epsilon_S
\end{equation}
with certain exponents $x_1$ and $x_2$. Scale invariance then implies that
\begin{equation}
\label{RawScaling}
\rho_S(\epsilon_A,\epsilon_S,t)=\Lambda^{-x_1} \, \rho_S(\Lambda \epsilon_A,\,\Lambda^{x_2} \epsilon_S,\Lambda^{-\nu_\parallel}t).
\end{equation}
Choosing again $\Lambda=t^{1/\nu_\parallel}$ we are led to the scaling form
\begin{equation}
\label{ScalingForm}
\rho_S(\epsilon_A,\epsilon_S,t)=t^{-x_1/\nu_\parallel}\,F_S(\epsilon_A\,t^{1/\nu_\parallel},\, \epsilon_S\, t^{x_2/\nu_\parallel})
\end{equation}
where $F_S$ is another scaling function. However, as we will see
below, this simple scaling form is not capable to describe the whole
dynamical evolution of $\rho_s$, instead it can be used only partially
either in the regimes I,II or in II,III, in each case with a different
set of exponents $x_1$ and $x_2$.  Furthermore, sufficiently close to
the curved critical line, the distance from the critical line becomes
a relevant parameter, and we will need a scaling form that depends on
two variables.

\subsection{Crossover between the regimes I and II}
%
In regimes I and II, the decay of $\rho_A$ is yet far from becoming stationary, and the decay of $\rho_S$ should be described by the same expression (\ref{ScalingForm}) everywhere near the multicritical point, and in particular on the two critical lines. 

\begin{figure}
\begin{center}
\includegraphics[width=88mm] {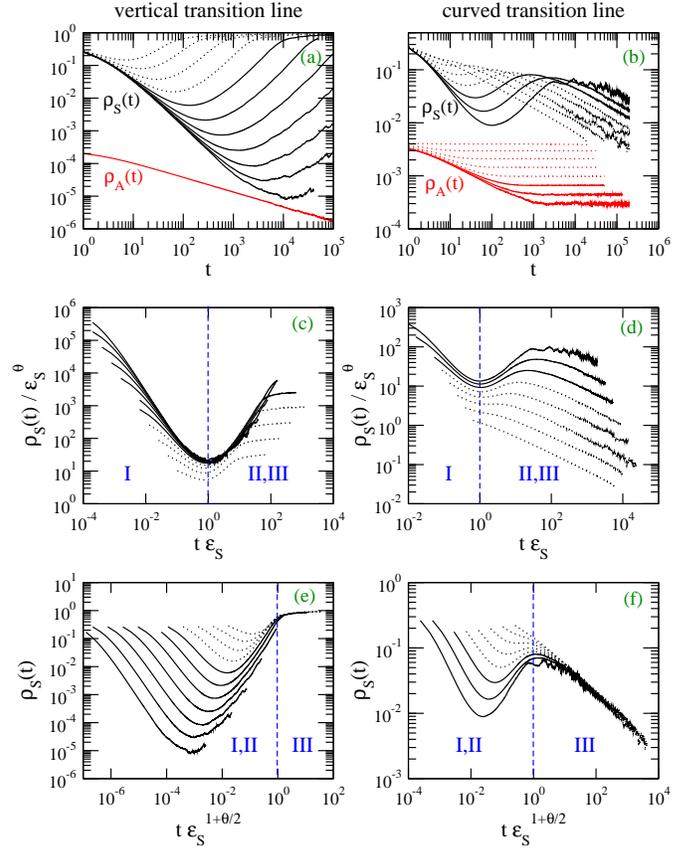}
\end{center}
\vglue -6mm
\caption{Simulations in 2+1 dimensions. The left column shows results obtained along the vertical transition line $\epsilon_A=0$ for $\epsilon_S=0.0001, 0.0002, \ldots ,0.1024$ while the right column displays similar results taken at the curved phase transition line for $\epsilon_A^*=0.0016,0.0032,\ldots,0.2048$. Data sets far away from the multicritical point ($\epsilon_S\geq0.0512$), where scaling is poor, are represented as dotted lines. \\
\hglue 4mm The panels are arranged in three rows. The first row shows the densities $\rho_S(t)$ and $\rho_A(t)$, the latter vertically shifted. In the second row the scaling form~(\ref{ScalingForm1}) is used to collapse the  curves in the dynamical regimes I and II (see text). Similarly, the third row shows data collapses based on the scaling form~(\ref{ScalingForm2}) which is valid in the regimes II and III. The respective crossovers are indicated by vertical dashed lines.
\label{fig:scaling}}
\end{figure}

In order to determine the exponents $x_1$ and $x_2$ in regimes I and II let us first consider the multicritical point $\epsilon_A=\epsilon_S=0$. Here the scaling form~(\ref{ScalingForm}) reduces to
$\rho_S(t)=t^{-x_1/\nu_\parallel}F_S(0,0)$ which -- compared  with Eq.~(\ref{Persistence})~--  gives
\begin{equation}
x_1=\theta \nu_\parallel \,.
\end{equation}
Moving up the vertical phase transition line, the DP process of species $A$ remains critical but species $S$ recovers after some time. It is natural to expect that the crossover from decay  (regime I) to subsequent growth (regime II)  takes place at a typical time scale
\begin{equation}
t_c^{(1)} \sim 1/\epsilon_S
\end{equation}
which is inversely proportional to the rate of spreading of species $S$. Scale invariance of the second argument of Eq.~(\ref{ScalingForm}) then implies that
\begin{equation}
x_2=\nu_\parallel.
\end{equation}
Therefore, we arrive at the scaling form
\begin{equation}
\label{ScalingForm1}
\rho_S(\epsilon_A,\epsilon_S,t)=t^{-\theta}\,F_S^{(1)}(\epsilon_A\,t^{1/\nu_\parallel},\, \epsilon_S t)\,.
\end{equation}
This scaling form can be verified along the vertical line by plotting
$\rho_S(t)\epsilon_S^\theta$ versus $t\epsilon_S$ for various values
of $\epsilon_S$ and checking for a data collapse.  As shown in the
second row of Fig.~\ref{fig:scaling}, this data collapse works nicely
at the crossover between the dynamical regime I to II (initial decay
and subsequent increase) while it clearly fails in regime III.  

The first argument of $F_S^{(1)}$ is in fact irrelevant, since the
density $\rho_S$ does not depend on it in regimes I and II. However,
it will be relevant in regime III, where $\rho_A$ becomes stationary
and determines the long-term behavior of $\rho_S$.
%

\subsection{Crossover between the regimes II and III}

The failure of a data collapse in regime III according to Eq.~(\ref{ScalingForm1}) suggests the presence of two different time scales in the system, namely, a crossover time scale
\begin{equation}
t_c^{(1)}\sim \epsilon_S^{-1}
\end{equation}
from where on offspring production becomes relevant, and a second time
scale $t_c^{(2)}$ from where on species $S$ feels the stationary
behavior of species $A$ which either induces a critical DP-like decay 
of species $S$ towards zero on the curved transition line or to an
asymptotically  stationary value of $\rho_S$ in the coexistence region 
and on the vertical line. To determine this second crossover time let
us first consider the vertical phase transition line, where species
$A$ evolves as a critical contact process. After the initial
persistence-like decay the remaining $S$ particles are the seeds of
$d$-dimensional
spheres growing linearly by offspring production essentially
unchallenged by species $A$. Therefore, in finite dimensions~$d$ the
density $\rho_S$ increases as $t^d$. On the vertical phase transition
line this growth continues until the density of species $S$ saturates
at the value $\rho_S \approx 1$, which defines the second crossover
time. Thus, matching initial decay and subsequent growth this
second crossover time has to scale as
\begin{equation}
t_c^{(2)} \sim \epsilon_S^{-1-\theta/d} \,.
\end{equation}
Therefore, in finite dimensions the two crossover times $t_c^{(1)}$ and $t_c^{(2)}$ scale \textit{differently} with respect to $\epsilon_S$ which explains the necessity of two separate scaling forms. 

The exponents $x_1$ and $x_2$ for the crossover from regime II to regime III can be determined as follows. On the one hand, the density $\rho_S$ saturates at the value $1$ for $t\to\infty$ on the vertical phase transition line, independent of $\epsilon_S$, implying that
\begin{equation}
x_1=0. 
\end{equation}
On the other hand, scale invariance of the argument of the scaling function in Eq.~(\ref{ScalingForm}) requires that
\begin{equation}
x_2=\frac{\nu_\parallel}{1+\theta/d}\,.
\end{equation}
Therefore, the resulting scaling form reads:
\begin{equation}
\label{ScalingForm2}
\rho_S(\epsilon_A,\epsilon_S,t)=F_S^{(2)}(\epsilon_A\,t^{1/\nu_\parallel},\, \epsilon_S t^{1/(1+\theta/d)})\,.
\end{equation}
This scaling form can be verified numerically by a data collapse along the vertical transition line, as demonstrated in Fig.~\ref{fig:scaling}e.
In Fig.~\ref{fig:scaling}f we demonstrate that this data collapse works not only at the vertical transition line but also along the curved phase transition line. Since the DP-like decay of $\rho_S$ is expected to set in when $\rho_A$ becomes stationary, i.e., at time $t\sim \epsilon_A^{-\nu_\parallel}$, it follows that the two time scales, namely, the DP correlation time $\xi_\parallel \sim \epsilon_A^{-\nu_\parallel}$ and the second crossover time $t_c^{(2)}\sim\epsilon_S^{-(1+\theta/d)}$ become identical on the curved critical line. 
More specifically, the phase transition line is expected to be characterized by a constant value of the scale-invariant combination $\epsilon_A^{-\nu_\parallel}/\epsilon_S^{-(1+\theta/d)}$, hence
\begin{equation}
\label{scalingrelation}
\epsilon_S^* \propto (\epsilon_A^*)^y\,,\qquad y=x_2=\frac{\nu_\parallel}{1+\theta/d}\,.
\end{equation}
Inserting the numerical estimates $\nu_\parallel=1.295(6)$ and $\theta=1.58(3)$ we obtain 
\begin{equation}
y=0.72(1)
\end{equation}
in fair agreement with the numerically extrapolated value in Sect.~\ref{PhaseTransitionLine}.

For large times, the expression (\ref{ScalingForm2}) for $\rho_S$ must approach an asymptotic value that is independent of time. We therefore obtain
\begin{equation}
\lim_{t\to\infty} F_S^{(2)}(z_1,z_2) = \frac{z_2}{z_1^\epsilon} \sim \frac{\epsilon_S}{\epsilon_A^y}\, .
\end{equation}
This implies that lines of constant density in the $\epsilon_S$-$\epsilon_A$-plane are given by constant ratios $\epsilon_S/\epsilon_A^y$. 

\begin{figure}
\begin{center}
\includegraphics*[width=85mm] {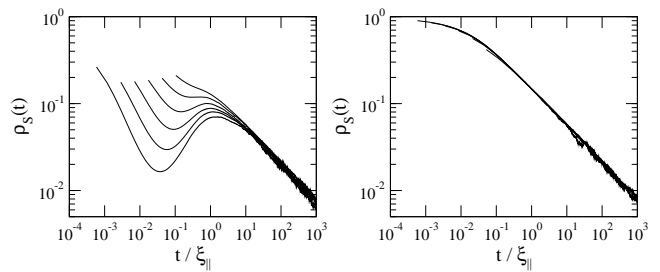}
\end{center}
\vglue -5mm
\caption{Left panel: Same data as in Fig.~\ref{fig:scaling}f, but with time
measured in units of $\xi_\parallel \sim \epsilon_A^{-\nu_\parallel}$ on the
horizontal axis.  Right panel: Simulation with the same parameters,
but with a stationary initial distribution of $A$ particles.
\label{stationary}}
\end{figure}

As we have mentioned, the scaling relation (\ref{scalingrelation}) implies
that the DP-like decay of species $S$ sets in only after its density
has increased to a considerable value. At this time species
$S$ has lost the memory of its past, and its dynamics is essentially
determined by species $A$. Therefore, the time evolution $\rho_S(t)$
is from then on identical to the one obtained starting the simulation
with a stationary $\rho_A$ and with a finite and random occupation
with $S$. This conclusion is supported by comparing the two
simulations, as shown in Fig.~\ref{stationary}. 
The range of the axes is the same in both panels, and one can
clearly see that for sufficiently large times the two types of curves
agree. This agreement of the large-time behavior for the two different
types of initial conditions occurs also in the mean-field theory, as
mentioned at the end of Section \ref{sec:mf}.

\subsection{Scaling near the curved transition line in regime III}

Finally let us study the asymptotic critical behavior of $\rho_S$ in the vicinity of the curved phase transition line. Close to the multicritical point  we have shown that this line can be parametrized by
\begin{equation}
\epsilon_S^*=a (\epsilon_A^*)^y\,,
\end{equation}
where $a$ is non-universal metric factor. On the curved line the density $\rho_S(t)$ decays asymptotically as $t^{-\delta}$, i.e., as in a critical DP process. Slightly above in the coexistence phase, $\rho_S(t)$ will eventually saturate at a value proportional to $(\Delta_S)^\beta$, where $\Delta_S=\epsilon_S-\epsilon_S^*$ denotes again the vertical distance from the line. This means that in addition to the previously discussed invariance under the transformation
\begin{equation}
\label{ScalingType1}
\epsilon_A\to\Lambda\epsilon_A,\quad
\epsilon_S\to\Lambda^y\epsilon_S,\quad
t\to\Lambda^{-\nu_\parallel}t
\end{equation}
the system \textit{simultaneously} has to be invariant under the DP-like scale transformation 
\begin{equation}
\label{ScalingType2}
\Delta_S\to\Omega\Delta_S,\quad
t\to\Omega^{-\nu_\parallel}t,\quad
\rho_S\to\Omega^\beta\rho_S
\end{equation}
with another scale factor $\Omega$ independent of $\Lambda$. The second type of scale invariance further constraints the form of $\rho_s(\epsilon_A,\epsilon_S,t)$, effectively reducing the number of independent arguments in the corresponding scaling form to~1. To see this it is convenient to switch from the parameters $(\epsilon_A,\epsilon_S)$ to the parameters $(\epsilon_A,\delta_S)$, where
\begin{equation}
\delta_S(\epsilon_A,\epsilon_S) = \frac{\Delta_S}{\epsilon_S^*(\epsilon_A)} = \frac{\epsilon_S}{a\epsilon_A^y}-1
\end{equation}
is the normalized distance from the critical line. In terms of these new parameters, the scaling form (\ref{ScalingForm2}) is expressed equivalently as
\begin{equation}
\label{NewScalingForm2}
\rho_S(\epsilon_A,\delta_S,t)=G_S(\epsilon_A\,t^{1/\nu_\parallel},\, \delta_S)\,,
\end{equation}
where $G_S(z_1,z_2)=F_S^{(2)}\bigl(z_1,\,az_1^y(1+z_2)\bigr)$. Obviously, the parameter $\delta_S$ is invariant under the first type of scale transformation~(\ref{ScalingType1}) while it scales as $\delta_S \to \Omega \delta_S$ under the second type of scale transformations (\ref{ScalingType2}). Applying the latter to the scaling form~(\ref{NewScalingForm2}) yields
\begin{equation}
\rho_s(\epsilon_A,\delta_S,t) = \Omega^{-\beta} \, G_S(\Omega^{-1}\epsilon_A\,t^{1/\nu_\parallel},\, \Omega \delta_S)\,.
\end{equation}
Setting $\Omega=\delta_S^{-1}$ one obtains
\begin{equation}
\label{FinalNewScalingForm}
\rho_S(\epsilon_A,\delta_S,t) = \delta_S^\beta \, \tilde{G_S}(\delta_S \,\epsilon_A \, t^{1/\nu_\parallel}).
\end{equation}
In fact, the argument $\delta_S \,\epsilon_A \, t^{1/\nu_\parallel}$ is a scale-invariant ratio under \textit{both} types of scale transformations, proving that the off-critical DP-process of species $A$ and the almost-critical DP-process of species $S$ can be described in terms of a unified scaling theory.
 We emphasize that this scaling form is valid \textit{only} in the temporal regime III close to the curved phase transition line, where $\delta_S \ll 1$. 

The scaling form (\ref{FinalNewScalingForm}) reproduces the expected DP behavior. Initially the system does not yet feel the influence of $\delta_S$, hence $\tilde{G_S}(z)\sim z^{-\beta}$ for small $z$. Later, for $\delta_S>0$, the initial DP-like decay of $\rho_S$ crosses over to a stationary value, meaning that for large arguments $\tilde{G_S}(z)$ becomes constant. Hence $\rho_S$ evolves as
\begin{equation}
\rho_S(\epsilon_A,\delta_S,t) \sim 
\left\{
\begin{array}{ll}
(t/\xi_\parallel)^{-\delta}
&
\mbox{ for } \,\,\, 1 \ll t/\xi_\parallel \ll \delta_S^{-\nu_\parallel}, \\[2mm]
\delta_S^\beta
&
\mbox{ for } \,\,\, t/\xi_\parallel \gg \delta_S^{-\nu_\parallel}\,,
\end{array}
\right.
\end{equation}
where $\xi_\parallel \sim \epsilon_A^{-\nu_\parallel}$ is the stationary correlation length of species $A$.

Exactly on the curved transition line, we have $\delta_S=0$, and
$\rho_S$ therefore decays forever. The unit time is set by
$\xi_\parallel$, which is the life time of the largest voids of the
$A$-process. Close to the curved transition line, a decay of $\rho_S$
with $t^{-\delta}$ occurs only if the time scale for saturation,
$\xi_\parallel \delta_S^{-\nu_\parallel}$ is larger than the crossover
time $t_c^{(2)}$, implying $\Delta_S \ll \epsilon_S$, or,
equivalently, $\delta_S \ll 1$. Otherwise, as we have already
mentioned, the scaling form (\ref{FinalNewScalingForm}) is not valid,
and saturation sets in directly after the growth in regime II, without
DP-like decay of $\rho_S$.

Finally, let us compare the results once more to mean-field theory. In
mean-field theory, the two crossover times scale in the same way,
making the distinction between two different scaling forms unnecessary
and removing the intermediate growth in regime II close to the
critical line, where the DP-like decay of $\rho_S$ becomes
visible. The values of the exponents are in mean-field theory
$\beta=y=\theta=\nu_\parallel=1$. The behavior in regime III near the
critical line that we found in mean-field theory resembles
qualitatively the one described in this section. In particular, the
proportionality factor between $\rho_S$ and $\Delta_S$ diverges at the
multicritical point, and so does the factor in front of the
$t^{-\delta}$ decay.

\section{Conclusions}
\label{sec:conclusions}

In this paper we have introduced a simple model of two competing
species $A$ and $S$ feeding on the same resources and living in
``patches'' represented by lattice sites. Species $A$ is characterized
by fast reproduction and can therefore displace species $S$ and
evolves independently according to the dynamic rules of a contact
process. Hence it belongs to the universality class of directed
percolation. Special $S$ has a lower reproduction rate and is
therefore restricted to live in those patches where $A$ is absent. In
contrast to species $A$, which can die out in a patch, patches
occupied by species $S$ do not become empty spontaneously.

In one spatial dimension, the phase diagram comprises two different phases where one of the species goes extinct, separated by a first-order transition line. In higher dimensions an additional mixed phase emerges, bounded by two continuous transition lines. This mixed phase is characterized by a non-trivial coexistence of the two species in the stationary state.\footnote{We note that by increasing the width of the one-dimensional system from 
one lattice site to several lattice sites (e.g. a strip of $2\times L$ sites), 
one can generate a modified version of the model where the two species 
can coexist even in one dimension, and where the phase transition of $S$ 
is also continuous.}

The curved phase transition line (except for the multicritical point) is found to belong to the DP universality class. This result may be surprising because $S$-individuals do not vanish spontaneously. On the other hand the result is also plausible since species $S$ `percolates' in the voids of the \textit{supercritical} DP of species $A$. Since the spatio-temporal arrangement of species $A$ involves only short-range correlations these voids are uncorrelated on large scales, which effectively leads to a DP transition. 

At the multicritical point the correlation lengths of the $A$-process
diverges, leading to a non-trivial scaling behavior. In this case,
starting with a mixed random initial conditions, $S$-occupied patches
mark those lattice sites that have been never visited by a species $A$
before. Therefore, the density $\rho_S$ decays in the same way as the
so-called local persistence probability in DP. For $\lambda_S>0$,
however, this decay crosses over to an increase of $\rho_S$ when the
reproduction of $S$ becomes visible. This increase continues until a
stationary state is reached in which the two species coexist if the
parameters are not too close to the curved phase transition
line. Close to the curved phase transition line, $\rho_S$ decreases
again after times long enough that $\rho_A$ has become stationary and
saturates eventually at a small nonzero value depending on the
distance to the critical line. We succeeded in introducing scaling
variables and scaling functions that characterize the behavior of
$\rho_S$ in the different regimes.



\begin{thebibliography}{99}

\bibitem{MarroDickman}  
J. Marro and R. Dickman, 
\textit{Nonequilibrium phase transitions in lattice models}, 
Cambridge University Press, Cambridge (1999).

\bibitem{Kinzel85}  
Kinzel W, Z. Phys. B {\bf 58}, 229 (1985).

\bibitem{Hinrichsen00}  
H. Hinrichsen, 
Adv. Phys. {\bf 49} 815 (2000).

\bibitem{OdorReview} 
G. {\'O}dor, 
Rev. Mod. Phys. {\bf 76}, 663 (2004).

\bibitem{Luebeck04}
S. L{\"u}beck, Int. J. Mod. Phys. B {\bf 18}, 3977 (2004).

\bibitem{Janssen81} 
H. K. Janssen, Z. Phys. B {\bf 42}, 151 (1981).

\bibitem{Grassberger82}
P. Grassberger, Z. Phys. B {\bf 47}, 365 (1982).

\bibitem{jan05} 
H-K Janssen and U.C. T\"auber, Ann. Phys. (NY) \textbf{315}, 147 (2005).

\bibitem{ZGB86}
R. M. Ziff, E. Gulari, and Y. Barshad", Phys. Rev. Lett. {\bf 56}, 2553 (1986).

\bibitem{GLB89}
G. Grinstein, Z. W. Lai, and D. A. Browne, Phys. Rev. A {\bf 40}, 4820 (1989).

\bibitem{JFD90}
I. Jensen, H. C. Fogedby, and R. Dickman, Phys. Rev. A {\bf 41}, 3411 (1990).

\bibitem{Jensen93}
I. Jensen, Phys. Rev. Lett. {\bf 70}, 1465 (1993).

\bibitem{MGDL96}
M. A. Mu{\~n}oz, G. Grinstein, R. Dickman, and R. Livi, Phys. Rev. Lett. {\bf 76}, 451 (1996).

\bibitem{tau98} 
U.C. T\"auber, M.J. Howard, and H. Hinrichsen, Phys. Rev. Lett. \textbf{80}, 2165 (1998).

\bibitem{gol99} Y.Y. Goldschmidt, H. Hinrichsen, M.J. Howard,
U.C. T\"auber, Phys. Rev. E \textbf{59}, 6381 (1999).

\bibitem{bas96} K.E. Bassler and D.A. Browne,
Phys. Rev. Lett. \textbf{77}, 4094 (1996).

\bibitem{can04} L. Canet, H. Chat\'e and B. Delamotte,
Phys. Rev. Lett. \textbf{92}, 255703 (2004).

\bibitem{spatialecology} D. Tilman, P. Karveia, Spatial Ecology: The
Role of Space in Popolation Dynamics and Inter-specific Interactions,
Princeton University Press, Princeton U.S.A. (1997)

\bibitem{pal92} S.C. Palmer and R.A. Norton, Biochemical systematics
and ecology \textbf{20}, 219 (1992).

\bibitem{scho98} I. Sch\"on, R.K. Butlin, H.I. Griffiths, K. Martens,
Proc. R. Soc. Lond. B \textbf{265}, 235 (1997).  

\bibitem{mul64} H.J. Muller, Mut. Res. \textbf{1}, 29 (1964).


\bibitem{Harris74}
T. E. Harris, Ann. Prob. {\bf 2}, 969 (1974).

\bibitem{JensenDickman93}
I. Jensen and R. Dickman, J. Stat. Phys. {\bf 71}, 89 (1993).

\bibitem{Dickman99}
R. Dickman, Phys. Rev. E {\bf 60}, 2441 (1999).

\bibitem{LuebeckWillmann05}
S. L{\"u}beck and R. D. Willmann, Nuclear Physics B 718, 341 (2005);
S. L{\"u}beck, Int. J. Mod. Phys. B {\bf 18}, 3977 (2004).

\bibitem{HinrichsenKoduvely98}
H. Hinrichsen and H. M. Koduvely, Eur. Phys. J. B {\bf 5}, 257 (1998).



\end{thebibliography}
\end{document}